\documentstyle[multicol,prl,aps,epsf]{revtex}
\begin{document}
\draft
\title{Vortex Response and Critical Fields observed via rf penetration depth
measurements on the superconductor ${\rm {\bf {YNi_2B_2C}}}$}
\author{S.~Oxx${}$, D.~P.~Choudhury${}$, Balam~A. Willemsen${}$, H. Srikanth${}$,
and S.~Sridhar${}^{*}$}
\address{Physics Department, Northeastern University, 360 Huntington Ave.,\\
Boston, MA 02115}
\author{B.~K.~Cho${}$ and P.~C. Canfield${}$}
\address{Ames Laboratory, Iowa State University, Ames, IA 50011}
\date{Submitted to Physica C February 19, 1996}
\maketitle

\begin{abstract}
Measurements of the rf penetration depth $\lambda (T,H,\theta )$ are used to
study the superconducting order parameter, vortex dynamics in the mixed
state and delineate critical fields in the borocarbide superconductor ${\rm %
YNi_2B_2C}$. The lower critical field has an anomalous $T$ dependence, $%
H_{c1}(T)=1.12\,\left[ 1-(T/T_c)\right] {\rm ~kOe}$, which is however
consistent with independent superfluid density measurements at microwave
frequencies. The vortex response is dominated by viscous flux flow,
indicative of extremely weak pinning, and is parametrized by a field scale $%
H_{c2,\,{\rm eff}}$. The angular dependence of the vortex contribution $%
\lambda (\theta )$ is in good agreement with the Coffey-Clem model.
Structure is seen in the depairing transition in the vicinity of the upper
critical field, with the existence of well-defined critical fields $H_{c2a}$%
, $H_{c2b}$ and $H_{c2c}$, with the vortex field scale $H_{c2,\,{\rm eff}}$
closest to $H_{c2b}$. Overall the measurements indicate that ${{\rm {%
YNi_2B_2C}}}$ has a rich and unusual field dependence of its transport
parameters.
\end{abstract}

\pacs{\sc{Keywords: Borocarbide Superconductors, Penetration Depth, Lower Critical
Field, Upper Critical Field, Flux Pinning}}

\begin{multicols}{2}

The borocarbide superconductors LnNi$_2$B$_2$C where Ln = \{Y, Lu, Tm, Er,
Ho and Dy\} \cite{RNagarajan94a,RJCava94a,BKCho95a}
are a recently discovered family of superconductors with elevated
transition temperatures. While there have been numerous studies of these
superconductors using traditional probes such as magnetization, specific
heat, tunneling, etc. there have been relatively few which directly probe
the order parameter and dynamics of vortices

In this paper we report on studies of the radio frequency (rf) penetration
depth $\lambda (T,H,\theta)$ of single crystal ${\rm YNi_2B_2C}$ ($T_c=15.5%
{\rm ~K}$) which were carried out in a highly sensitive rf tunnel diode
oscillator set-up. The sample is placed in an inductive coil which forms
part of a tank circuit, which is self-resonant typically at $3{\rm ~MHz}$
and is driven by a tunnel diode circuit. Changes in the penetration depth or
skin depth of the sample, caused by varying temperature, $T$, applied dc
magnetic field, $H$, or angle between $H$ and the $ab$ plane, $\theta$, lead
to changes in the coil inductance. These are then detected as changes in
resonant frequency, and converted to changes in $\lambda$ via the relation $%
\delta \lambda (T,H,\theta)\equiv-g\,[f(T,H,\theta)-f_0(T,H)]$, where $g$ is
a geometrical factor set by the sample dimensions and $f_0$ refers to the
resonant frequency without the sample. The very high stability of the
circuit, typically $1{\rm ~Hz}$ in $3{\rm ~MHz}$, leads to a very high
resolution of the order of a few \AA\@. The technique has been extensively
validated through precise measurements in the cuprate superconductors of the
non-linear Meissner effect and of vortex parameters such as $H_{c1}$ and
pinning force constants\cite{SSridhar89a,DHWu90a}. In one experimental
set-up the sample can be oriented with $H\parallel \hat c$ ($%
\theta=90^{\circ}$) or $H\perp \hat c$ ($\theta=0^{\circ}$), and fields up
to $7{\rm ~T}$ can be applied. In another set-up the angle $\theta$ can be
varied continuously between $0^{\circ}$ and $90^{\circ}$ with an angular
accuracy of $0.2^{\circ} $. Here the maximum applied field is $6.4{\rm ~kOe}$%
. In all cases the rf field $H_{{\rm rf}}\parallel ab$ and $H_{{\rm rf}%
}\perp H$.

The single crystals of ${\rm YNi_2B_2C}$ used for these measurements were
grown using the Ames Lab ${\rm Ni_2B}$ flux growth method\cite
{MXu94a,BKCho95b}. Crystals grown via this method are in the form of plates
with the crystallographic $\hat{c}$ axis perpendicular to the surface of the
plate. The crystal used for these measurements has approximate dimensions $%
1.4\times 1.2\times 0.2{\rm ~mm^3}$. 

\section{Lower Critical Field: H$_{c1}$}

\begin{figure}
  \narrowtext
  \begin{center}
    \epsfclipon
    \epsfxsize 0.45\textwidth
    \epsfbox[13 13 780 599]{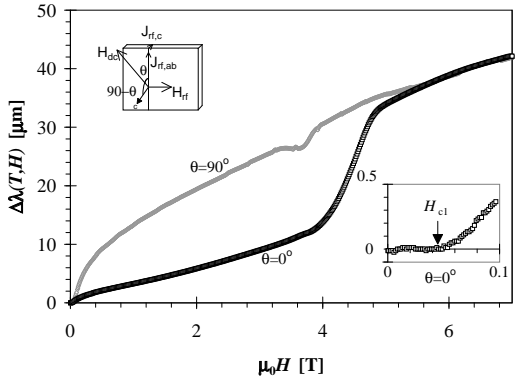}
  \end{center}
  \caption{$\Delta\lambda$ {\it vs.}\/ $H$ at $T = 8{\rm{~K}}$ for
    ${\rm{YNi_2B_2C}}$. (Inset) Expanded plot of low field
    portion, showing onset at $H_{c1}$. A schematic of the
    field/sample orientation is also presented as an inset}
  \label{Fig1}
\end{figure}

Typical results for $\Delta \lambda (T,H,\theta )\equiv \delta \lambda
(T,H,\theta )-\delta \lambda (T,H=0)$ {\it vs.}\/ $H$ are shown in Fig.~\ref
{Fig1}, for both $\theta =0^{\circ }$ and $90^{\circ }$. The low field
portion is further shown in detail in the inset, which reveals essentially
no change until a critical field $H_{c1}$ is reached, above which $\lambda $
increases rapidly. This critical field $H_{c1}$ represents the field at
which flux first enters the sample and is governed by both the Meissner
state of the superconductor and an effective geometric barrier at the
surface. We are justified in calling it the lower critical field as
demagnetization corrections are negligible in the case where $H
\parallel ab$. Note that the presence of surface barriers would lead to
hysteresis in the field dependence of $\Delta \lambda (T,H,\theta )$ at the
onset of flux entry. In our data, the observed hysteresis is minimal and is
within the experimental resolution. It is also to be noted that surface
barrier effects would scale the field values but do not affect the
temperature dependence. 

Notice that unlike magnetization measurements, where $H_{c1}$ is
deduced from deviations from linearity in the $M(H)$ curve, the signature in
our data at $H_{c1}$ is very sharp. This is because the experiment
effectively measures the flux density $B(H)$ (as will become evident later).
Hence, the signature at $H_{c1}$ is a change from zero, and not a change of
slope as in $M(H)$ data.

\begin{figure}
  \narrowtext
  \begin{center}
    \epsfclipon
    \epsfxsize 0.45\textwidth
    \epsfbox[13 13 780 599]{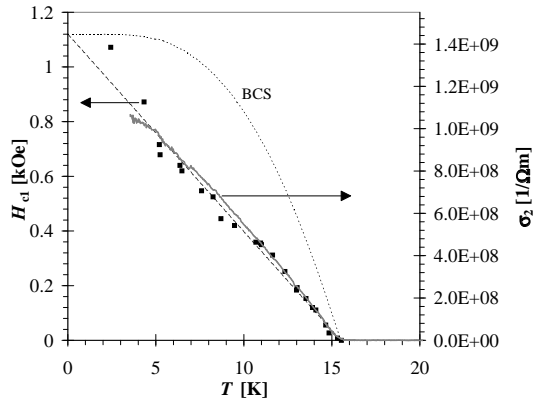}
  \end{center}
  \caption{$H_{c1}$ {\it vs.}\/ $T$. $\sigma_2$ {\it vs.}\/ $T$ from 
    Ref.~\protect\cite{TJacobs95b} is shown for comparison. The dotted
    line represents a BCS temperature dependence of $H_{c1}$ while the
    dashed line is a $1-t$ dependence obtained from a least squares
    fit to the data.}
  \label{Fig2}
\end{figure}

The $T$ dependence of $H_{c1}(T)$ for ${\rm YNi_2B_2C}$ is shown in Fig.~\ref
{Fig2}. This essentially has a linear behaviour from $T_c$ to $2{\rm ~K}$,
well represented by $H_{c1}(T)=1.12(1-t){\rm ~kOe}$, where $t=T/T_c$. This
is unlike that seen in conventional low $T_c$ superconductors and even
differs from the behavior of ${{\rm YBa_2Cu_3O_{7-\delta }}}$. However this
anomalous behavior finds support in measurements on a related quantity, the
superfluid density, measured via the condensate contribution to the complex
conductivity, $\sigma _2$, which we have measured elsewhere \cite{TJacobs95b}
using a $10{\rm ~GHz}$ ${\rm Nb}$ cavity. There, $\sigma _2$ was shown to
have an anomalous temperature dependence well represented by $\sigma
_2(T)=10^9(1-t){{\rm ~(\Omega m)^{-1}}}$ (also shown in Fig.~\ref{Fig2}).
Now $\sigma _2=1/\mu _0\omega \lambda ^2$, and since $H_{c1}\sim (\Phi
_0/4\pi \lambda ^2)\left[ \ln (\kappa )+0.5\right] $, there is a good
correlation between the linear temperature dependence of the $H_{c1}$ and $%
\sigma _2$ measurements. (Note that for clean type-{II} superconductors, $%
\kappa $ varies as $1/(1+t^2)$ and thus $\ln (\kappa )$ is expected to have
a weak $T$ dependence\cite{RDParks69}).

\section{Vortex Dynamics and Depairing}

Above $H_{c1}$, $\lambda (T,H)$ increases strongly in the $\theta =0^{\circ
} $ case, due to the increasing $B$ in the sample. A similar increase also
occurs in the $\theta =90^{\circ }$ data, although here the effective $%
H_{c1}\rightarrow 0$ due to demagnetization. This field dependence $\lambda
(H)$ for $H>H_{c1}$ arises from two effects of the magnetic field on the
superconductor:\ depairing of condensed electrons (which dominates when the
rf current $J_{{\rm rf}}\parallel H$ ) and dynamic vortex response (which
dominates when $J_{{\rm rf}}\perp H$ ). 
This is followed by a sharp increase starting and terminating at field
values which we call $H_{c2a}$ and $H_{c2b} $ respectively. Above $H_{c2b}$, 
$\lambda$ increases almost linearly with $H$, which is consistent with the
normal state showing a strong positive magnetoresistance. There is an
additional field $H_{c2c}$ just above $H_{c2b}$ which can be identified as
the point where the $\theta=0^{\circ }$ and $90^{\circ }$ curves meet as
seen in Fig. 1.

We have observed a similarly rich structure in the transition region of the $%
{\rm Ln}=\{{\rm Er},{\rm Ho}$ and ${\rm {Dy}}\}$ compounds \cite{Structure}. 
The presence of multiple structure in the
transition region, although interesting in its own right, makes the
identification of a single upper critical field $H_{c2}$ somewhat ambiguous.
This difficulty is also compounded by the linear-$H$ magnetoresistance in
the normal state. All three field scales ($H_{c2a}$, $H_{c2b}$ and $H_{c2c}$%
) agree quantitatively with the $H_{c2}$ values reported in the literature%
\cite{MXu94a}. While the exact origin of this structure is not yet precisely
known, its presence in all the borocarbide superconductors we have studied
seems to suggest an intrinsic phenomenon and is not likely to be due to
spurious chemical phases, as will be presented in a forthcoming publication 
\cite{Structure}.

In this paper, we focus on the general nature of the depairing and vortex
contributions to the rf penetration depth. To test the validity of the
standard Coffey-Clem model for the depairing and vortex response in this
system, an effective upper critical field $H_{c2}$ coinciding with $H_{c2b}$
is assumed for subsequent analysis. It is to be pointed out that our choice
of $H_{c2b}$ as the effective upper critical field is not arbitrary and is
based on other factors like the validity of the Bardeen-Stephen
approximation \cite{JBardeen65a} which relates the upper critical field to
the flux flow resistivity. A discussion on flux viscosity presented later in
the paper will clarify this issue further.

\section{Detailed Comparison To Theory}

We now turn to an understanding of the field dependence of $\lambda
(T,H,\theta )$. The principal causes for this field dependence are the
depairing effect of $B(H)$ on the condensate and the dynamic response of the
vortices to the Lorentz force induced by the rf current. These contributions
have been treated self-consistently by Coffey and Clem, leading to the
following expression for the complex penetration depth which includes the
full angular dependence: \cite{MWCoffey92a} 
\begin{equation}
{\tilde{\lambda}}\equiv \lambda -i\frac{R_s}{\mu _0\omega }=\left[ \frac{%
\lambda _L^2+\Phi _0B\sin ^2\theta /\mu _0(\kappa _p-i\omega \eta )}{%
1+i(2\lambda _L^2/\delta _{nf}^2)}\right] ^{1/2}  \label{Eq:CC}
\end{equation}
where $\lambda _L(T,B)$ and $\delta _{nf}(T,B)$ are the condensate
penetration depth and the normal-fluid skin depth in the superconducting
state, and $\kappa _p(T)$ and $\eta (T)$ are the vortex pinning force
constant and viscosity respectively. The field-dependence of the condensate
background is represented by $\lambda _L^2(T,B)=\lambda _{L0}^2\,f(T,B)$ and 
$\delta _{nf}^2(T,B)=\delta _n^2(T,B)\,\left[ 1-f(T,B)\right] $, where for
instance $f(T,B)=[1-(T/T_c)^4]\left[ 1-(B/B_{c2})\right] $ in a two-fluid
model. For $\theta =0^{\circ }$, the rf currents in the $ab$ plane are
parallel to the vortices and hence the Lorentz force is zero for those
vortices at the sample faces. (A schematic representation of the relative
orientations of $H_{{\rm rf}}$ and the rf current, $J_{{\rm rf}}$, with
respect to $H$ is shown as an inset in Fig.~\ref{Fig1}). Thus the vortex
response is mostly absent in the $\theta =0^{\circ }$ case, except for a
small contribution of order $O(d/W)\sim 0.1$ from the edges, where $d$ and $%
W $ are the sample thickness and width respectively. This can be
incorporated phenomenologically in terms of an aspect ratio $r\equiv \left[
1+\left( d/W\right) ^{-1}\right] ^{-1}$ which mixes the vortex contribution (%
$\theta =90^{\circ }$) to the force-free ($\theta =0^{\circ }$) term as $%
\sqrt{{\tilde{\lambda}}_{||}^2+r{\tilde{\lambda}}_{\perp }^2}$. The angular
dependence provides a nice way of separating out the depairing and the
vortex contributions to the penetration depth.

\begin{figure}[htbp]
  \narrowtext
  \begin{center}
    \epsfclipon
    \epsfxsize 0.45\textwidth
    \epsfbox[13 13 780 599]{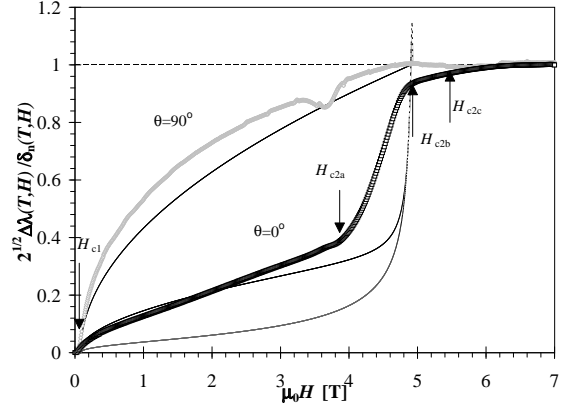}  
  \end{center}
  \caption{$2^{1/2}\Delta\lambda(T,H)/\delta_n(T,H)$ {\it vs.}\/ $H$ at
    $T=8{\rm{~K}}$ and $\theta=0^{\circ}$ and $90^{\circ}$. The
    solid lines represent theoretical calculations based on
    Eq.~\protect\ref{Eq:CC} for the two orientations presented,
    including finite aspect ratio effects for the $\theta=0^{\circ}$
    case with $r=0.11$. The dotted line ignores the effects of finite
    aspect ratio. The dashed horizontal line represents the normal
    state skin depth.}
  \label{Fig:Norm}
\end{figure}

In Fig.~\ref{Fig:Norm} we present the same data as in Fig.~\ref{Fig1}
normalized with respect to $\delta _n(T,H)/\sqrt{2}$. Motivated by the
observation that the field dependence of $\delta _n(T<T_c,H>H_{c2})$ follows
that of $\delta _n(T>T_c,H)$ we have used $\delta _n(T,H)\approx \delta
_n(T>T_c,H)$ for this normalization. Theoretical calculations based on Eq.~%
\ref{Eq:CC}, taking into account the small aspect ratio as discussed earlier
and also assuming negligible pinning (i.e. $\kappa _p$ $\rightarrow 0$), are
shown in Fig.~\ref{Fig:Norm} along with the data. For the vortex
contribution, at fixed frequency $f$ the response is pinning dominated if $%
f<f_c$ and viscosity dominated if $f>f_c$, where $f_c=\kappa _p/2\pi \eta $.
It is quite striking that pinning is negligible even at $3{\rm ~MHz}$, which
would effectively imply that flux is in free flow as soon as vortices enter
the sample above $H_{c1}$. This is in contrast to the estimated $f_c$ values
of $20{\rm ~GHz}$ in ${\ {\rm YBa_2Cu_3O_{7-\delta }}}$\cite{BAWillemsen94b}
and $\sim 100{\ {\rm ~MHz}}$ in low $T_c$ materials \cite{JIGittleman64a}.
For $\theta =0^{\circ }$, the low field variation of $\lambda $ is
adequately reproduced by the calculation, confirming the presence of a
finite vortex contribution consistent with the aspect ratio independently
determined from the sample dimensions.

The $\theta =90^{\circ }$ data shows a dip at a field below the effective $%
H_{c2}$. This can result from sudden changes in pinning or scattering which
are parametrized by the terms $\kappa _p$ and $\eta $ in Eq.~\ref{Eq:CC}. A
rise in $\kappa _p$ at a threshold field can be associated with the ``peak
effect'' commonly seen in type-{II} superconductors with weak pinning. The
signature of the peak effect is a peak in critical current density and would
show up as a dip in the rf susceptibility, which is what we are effectively
measuring in the $\theta =90^{\circ }$ case. There is a general consensus
that the peak effect arises due to enhanced pinning as a result of a rapidly
decreasing rigidity of the vortex lattice just below $H_{c2}\ $\cite
{AILarkin79a}. However, there is still some controversy over whether it is a
softening of the tilt modulus ($C_{44}$) or shear modulus ($C_{66}$) that
leads to this decrease in rigidity \cite{EHBrandt77a,EHBrandt86a,XSLing95a}.
An alternate possibility is that this dip feature is due to an anomalous
field dependence of $\eta $ arising as a result of unusual quasiparticle
scattering in the superconducting state. Since the effect is much stronger
in the vortex response rather than the depairing, this is an indication that
the scattering due to bound quasiparticle states in the normal vortex cores
could also be responsible for the observed dip in the $\theta =90^{\circ }$
data below the effective $H_{c2}$.

When the vortex term dominates as at $\theta =90^{\circ }$, the $\lambda
(T,H)$ data can be used to extract the pinning forces and viscosity as we
have done elsewhere in both single crystals and films of ${\ {\rm %
YBa_2Cu_3O_{7-\delta }}}$\cite{DHWu90a,BAWillemsen94b}. A Bardeen-Stephen
estimate of the viscosity can be made using $\eta _{{\rm BS}}(T,B)=\Phi
_0B_{c2}(T)/\rho _n(T,B)$, which yields $\eta _{{\rm BS}}(0,0)\sim 1\times
10^{-6}{\rm ~Ns/m^2}$ for ${\rm YNi_2B_2C}$, compared to values of $\sim
10^{-8}{\rm ~Ns/m^2}$ for ${\rm Nb}$.

\begin{figure}[htbp]
  \narrowtext
  \begin{center}
    \epsfclipon
    \epsfxsize 0.45\textwidth
    \epsfbox[13 13 780 599]{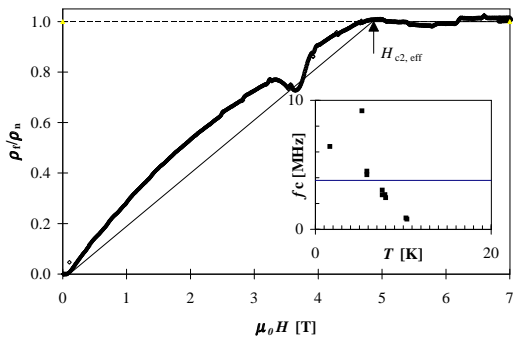}  
  \end{center}
  \caption{Flux flow resistivity $\rho_f/\rho_n$ obtained from the 
    $\theta=90^{\circ}$ data in Fig.~\protect\ref{Fig:Norm}. The solid
    line indicates the expected Bardeen-Stephen behavior. The dashed
    horizontal line indicates the normal state resistivity. (inset)
    $f_c=\kappa_p/\eta$ {\it vs.}\/ $T$ extracted from the relative
    peak height is presented as an inset. The horizontal line in the
    inset indicates the measurement frequency.}
  \label{Fig:VE}
\end{figure}

The data in Fig.~\ref{Fig:Norm} can also be presented as the normalized flux
flow resistivity $\rho _f/\rho _n\equiv 2\lambda ^2(T,B)/\delta _n^2(T,B)$
which is presented as a function of $H$ for $T/T_c=0.5$ in Fig.~\ref{Fig:VE}%
. The flux flow resistivity approaches the normal state almost linearly with
a slope $1/B_{c2,\,{\rm eff}}$ $(\sim 1/B_{c2})$ consistent with the
observations of Gittleman and Rosenblum in conventional superconductors\cite
{JIGittleman64a}. Overall the response appears to be well described by Eq.~%
\ref{Eq:CC} assuming $\eta =\eta _{{\rm BS}}(T,B)$ and $\kappa _p\rightarrow
0$, consistent with magnetization experiments which have also shown evidence
for extremely weak pinning\cite{MXu95p}. The differences between the
observed response and theoretical Bardeen-Stephen response (indicated by the
solid line in Fig.~\ref{Fig:VE}) 
can be attributed to an underestimate of $\rho _n(T,B)$ in the
superconducting state as well as some variation of $\kappa _p(T)$. If the
dip observed in the $\theta =90^{\circ }$ data is associated with a change
in pinning, we can estimate the change in $\kappa _p$ in the peak region
from the relative depth of the dip since $\eta _{{\rm eff}}\sim \eta /({%
1+\kappa _p/\omega \eta })$. A plot of $f_{c,\,{\rm peak}}=\kappa _p/2\pi
\eta (T)$ extracted in this form is shown in the inset to Fig.~\ref{Fig:VE}.

\section{Angular dependence}

\begin{figure}[htbp]
  \narrowtext
  \begin{center}
    \epsfclipon
    \epsfxsize 0.45\textwidth
    \epsfbox[13 13 780 599]{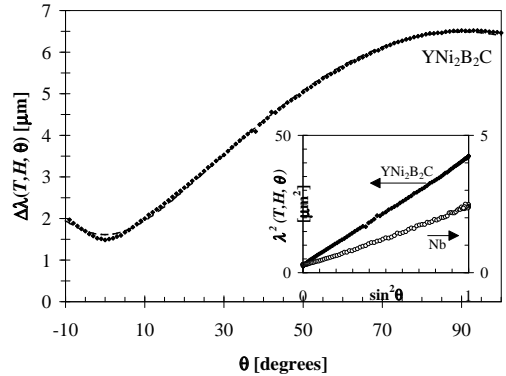}
  \end{center}
  \caption{$\Delta\lambda$ {\it vs.}\/ $\theta$ at $T=8\rm{~K}$ and 
    $H=1{\rm{~kOe}}$. The line represents a least squares fit to Eq.~\protect\ref{Eq:CC}. (Inset) $\lambda^2$ {\it vs.}\/
    $\sin^2\theta$ for the same data for comparison with
    Eq.~\protect\ref{Eq:CC}. A plot for polycrystalline
    ${\rm{Nb}}$ at $T=7{\rm{~K}}$ and $H=1.5{\rm{~kOe}}$
    is also shown.}
  \label{Fig:Ang}
\end{figure}

The full angular dependence of $\lambda $ for $H_{c1}<H<H_{c2a}$ is shown in
Fig.~\ref{Fig:Ang} along with the fit generated using Eq.~\ref{Eq:CC}. The
inset shows $\lambda ^2$ plotted against $\sin ^2\theta $ for the same data.
A similar plot for a poly-crystalline ${\rm Nb}$ sample of comparable
dimensions is also shown in the panel. The excellent agreement between the
experimental data and the Coffey-Clem expression of Eq.~\ref{Eq:CC} is
obvious from the good fit and also the linear dependence of $\lambda ^2$
with $\sin ^2\theta $. The slopes are vastly different for ${\rm YNi_2B_2C}$
and ${\rm Nb}$ as the response is pinning dominated in ${\rm Nb}$ whereas
pinning is almost irrelevant in ${\rm YNi_2B_2C}$ at $3{\rm ~MHz}$. The
Coffey-Clem calculation as presented in Eq.~\ref{Eq:CC} assumes an isotropic 
$H_{c2}$ case and from the good fit with the ${\rm YNi_2B_2C}$ data one can
infer that the effective $H_{c2}$ is generally isotropic in this system\cite
{MXu94a}, although a slight deviation of the data from the model at low
angles might indicate the presence of some degree of anisotropy in these
layered systems. Elsewhere we have observed unusual angular dependence and
anisotropy effects in ${\rm HoNi_2B_2C}$ related to magnetic order and
details about this will be given in a separate publication.

\begin{figure}[htbp]
  \narrowtext
  \begin{center}
    \epsfclipon
    \epsfxsize 0.45\textwidth
    \epsfbox[13 13 780 599]{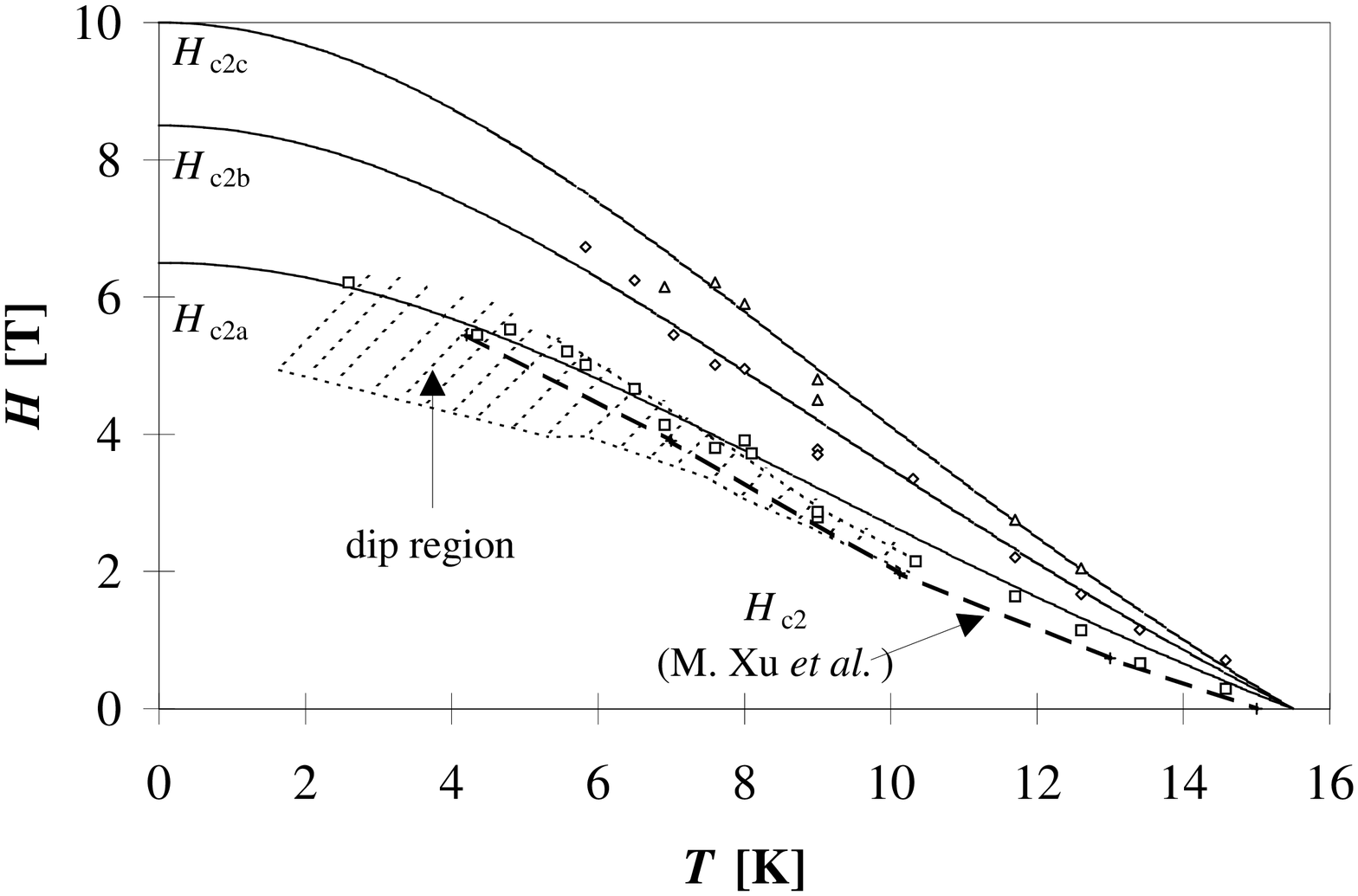}  
  \end{center}
  \caption{$H_{c2a}$, $H_{c2b}$ and $H_{c2c}$ {\it vs.}\/ $T$ obtained
    from the $\theta=0^{\circ}$ data.  Lines representing a
    $(1-t^2)/(1+t^2)$ dependence are overlaid.  $H_{c2}$ values from
    Ref.~\protect\cite{MXu94a} are indicated by the thick dashed line. The
    region where the dip is observed in the $\theta=90^{\circ}$ data
    is indicated by the dotted hatch lines.}
  \label{Fig:HT}
\end{figure}

The magnitude and $T$ dependence of the three features in the vicinity of $%
H_{c2}$ for ${\rm YNi_2B_2C}$ ($H_{c2a}$, $H_{c2b}$ and $H_{c2c}$) are shown
in Fig.~\ref{Fig:HT} are in good agreement with those determined by
magnetization measurements \cite{MXu94a} and the overall temperature
dependence is fit reasonably well by a $(1-t^2)/(1+t^2)$ form. Two aspects
make the observed onset field $H_{c2a}$ extremely interesting: $H_{c2a}$ is
located at about the same field where the dip is seen in the $(\theta
=90^{\circ })$ data and occurs further from the effective upper critical
field than would otherwise be expected. The theoretical calculation fails to
reproduce the finite slope of the transition, which would imply an unusual
field dependence of $\lambda (B)$. Looking at it in a different way, the
simple variation of the superfluid density ($=\lambda^2_0/\lambda ^2$) as ($1-B/B_{c2}$%
) within the framework of a two-fluid model breaks down and one has to
assume a stronger reduction in the superfluid density as $B\rightarrow B_{c2}
$. Note that earlier studies on ${\rm YNi_2B_2C}$\cite{TJacobs95b} using a
superconducting ${\rm Nb}$ cavity at $10{\rm ~GHz}$ also indicated
disagreement with BCS or two-fluid models in the $T$ dependence of the
surface resistance and penetration depth, as displayed in Fig.~\ref{Fig2} in
terms of $\sigma _2(T)$. This may indicate the existence of scattering of
unknown origin in the superconducting state. We believe the origin of $%
H_{c2a}$ could result from this anomalous scattering or vortex lattice
softening/melting or a combination of both these effects. Whatever the
cause, the underlying dependence of $\Delta \lambda (T,H,\theta =0)$ in the
vicinity of $H_{c2a}$ is masked to some extent by the mixing of vortex and
depairing terms in the $(\theta =0^{\circ })$ data.

In conclusion, the rf measurements presented here reveal a rich variety of
electrodynamic response in the superconducting state of the non-magnetic
borocarbide superconductor ${\rm YNi_2B_2C}$. In particular, $\lambda
(T,H=0) $ and $H_{c1}$ both have anomalous non BCS $T$ dependences. The
experiments have very high sensitivity which enable observation of subtle
effects on the superconducting order parameter and the vortex lattice due to
the applied field. A unique way of studying the vortex response and
field-induced depairing individually by variation of $\theta$ is presented.
The angular dependence of $\lambda (T,H,\theta ) $ is well described by the
Coffey-Clem theory with an isotropic $H_{c2}$. A dip is observed in the
vortex response which can be related to a change in pinning (and thus to the
``peak effect'') or alternatively to anomalous scattering in the mixed
state. Pinning is found to be extremely weak, and consequently, the crossover
frequency is expected to be orders of magnitude lower than that of most
conventional and high-$T_c$ superconductors. Structure is observed in the depairing
transition from the
superconducting to normal state, which is also found to have an anomalous $H$
dependence. Thus although ${\rm YNi_2B_2C}$ is in many ways a conventional
high-$\kappa $ type-{II} superconductor, the high frequency measurements
appear to indicate many unconventional properties of the Meissner and mixed
states.

Work at Northeastern was supported by grant NSF-DMR-9223850. Ames Laboratory
is operated for the U.~S. Department of Energy by Iowa State University
under Contract No.\ W-7405-Eng-82. Work at Ames was supported by the
Director for Energy Research, Office of Basic Energy Sciences.


\begin{references}
\bibitem[*]{E-mail}{E-mail: srinivas@neu.edu}

\bibitem{RNagarajan94a}
R.~Nagarajan, C.~Mazumdar, Z.~Hossain, S.~K. Dhar, K.~V. Gopalakrishnan, L.~C.
  Gupta, C.~Godart, B.~D. Padalia, and R.~Vijayaraghavan,
\newblock Phys. Rev. Lett. {\bf 72}, 274 (1994).

\bibitem{RJCava94a}
R.~J. Cava, H.~Takagi, B.~Battlog, H.~W. Zandbergen, J.~J. Krajewski, W.~F.~P.
  jr., R.~B. van Dover, R.~F. Felder, T.~Siegrist, K.~Mizuhaski, J.~O. Lee,
  H.~Eisaki, S.~A. Carter, and S.~Uchida,
\newblock Nature {\bf 367}, 146 (1994).

\bibitem{BKCho95a}
B.~K. Cho, P.~C. Canfield, L.~L. Miller, and D.~C. Johnston,
\newblock Phys. Rev. B {\bf 52}, R3844 (1995).

\bibitem{SSridhar89a}
S.~Sridhar, D.-H. Wu, and W.~L. Kennedy,
\newblock Phys. Rev. Lett. {\bf 63}, 1873 (1989).

\bibitem{DHWu90a}
D.-H. Wu and S.~Sridhar,
\newblock Phys. Rev. Lett. {\bf 65}, 2074 (1990).

\bibitem{MXu94a}
M.~Xu, P.~C. Canfield, J.~E. Ostenson, D.~K. Finnemore, B.~K. Cho, Z.~R. Want,
  D.~C. Johnston, and D.~E. Farrell,
\newblock Physica C {\bf 227}, 321 (1994).

\bibitem{BKCho95b}
B.~K. Cho, M.~Xu, P.~C. Canfield, L.~L. Miller, and D.~C. Johnston,
\newblock Phys. Rev. B {\bf 52}, 3676 (1995).

\bibitem{TJacobs95b}
T.~Jacobs, B.~A. Willemsen, S.~Sridhar, P.~C. Canifeld, and B.~K. Cho,
\newblock Phys. Rev. B {\bf 52}, R7022 (1995).

\bibitem{RDParks69}
R.~D. Parks, editor,
\newblock {\em Superconductivity},
\newblock Marcel Dekker, Inc., New York, 1969.

\bibitem{Structure}
S.~Sridhar et~al.,
\newblock in preparation.

\bibitem{JBardeen65a}
J.~Bardeen and M.~J. Stephen,
\newblock Phys. Rev. {\bf 140}, A1197 (1965).

\bibitem{MWCoffey92a}
M.~W. Coffey and J.~R. Clem,
\newblock Phys. Rev. B {\bf 45}, 9872 (1992).

\bibitem{BAWillemsen94b}
B.~A. Willemsen, J.~S. Derov, J.~H. Silva, and S.~Sridhar,
\newblock Appl. Phys. Lett. {\bf 67}, 551 (1995).

\bibitem{JIGittleman64a}
J.~I. Gittleman and B.~Rosenblum,
\newblock Proc. IEEE , 1138 (1964).

\bibitem{AILarkin79a}
A.~I. Larkin and Y.~N. Ovchinnikov,
\newblock J. Low. Temp. Phys. {\bf 34}, 409 (1979).

\bibitem{EHBrandt77a}
E.~H. Brandt,
\newblock J. Low. Temp. Phys. {\bf 26}, 709 (1977).

\bibitem{EHBrandt86a}
E.~H. Brandt,
\newblock Phys. Rev. B {\bf 34}, 6514 (1986).

\bibitem{XSLing95a}
X.~Ling, C.~Tang, S.~Bhattycharya, and P.~M. Chaikin,
\newblock Peak effect in superconductors: Melting of larkin domains,
\newblock from cond-mat. 9504109

\bibitem{MXu95p}
M.~Xu and P.~C. Canfield,
\newblock private communication.

\end{references}

\end{multicols}

\end{document}